\documentclass[aps,amsmath,showpacs,amsfonts,10pt]{revtex4}
\usepackage{epsfig,graphicx}
\usepackage[english]{babel}
\usepackage{amsfonts}
\usepackage{amsmath}
\usepackage{latexsym}
\usepackage{graphics,bm}
\usepackage{natbib}
\usepackage{dcolumn}
\usepackage{bm}
\usepackage{rotating}

\begin{document}

\title{Opto-mechanical effects in superradiant light scattering by Bose-Einstein condensate in a cavity}

\author{Aranya B Bhattacherjee$^{1,2}$ }

\address{$^{1}$Department of Physics, ARSD College, University of Delhi (South Campus), New Delhi-110021, India}
\address{$^{2}$ Institut f\"{u}r Theoretische Physik, Freie Universit\"{a}t,Arnimallee 14, 14195, Berlin, Germany}

\begin{abstract}
We investigate the effects of a movable mirror (cantilever) of an optical cavity on the superradiant light scattering from a Bose-Einstein condensate (BEC) in an optical lattice. We show that the mirror motion has a dynamic dispersive effect on the cavity-pump detuning. Varying the intensity of the pump beam, one can switch between the pure superradiant regime and the Bragg scattering regime. The mechanical frequency of the mirror strongly influences the time interval between two Bragg peaks. We found that when the system is in the resolved side band regime for mirror cooling, the superradiant scattering is enhanced due to coherent energy transfer from the mechanical mirror mode to the cavity field mode.
\end{abstract}


\maketitle

\section{Introduction}

Bose-Einstein condensates, which are ensemble of atoms in a single quantum state with long coherence time, offers the possibility to study quantum optics in a completely new regime. The high degree of spatial and temporal coherence of a condensate can alter the interactions between atoms and light. In particular, superradiant Rayleigh scattering and matter-wave amplification caused by coherent center-of-mass motion of atoms in a condensate illuminated by a highly detuned laser were observed \citep{Inouye, Schneble, kozuma}. The matter-wave grating is formed as a result of the coherent superposition of different atomic states, similar to the one produced in Bragg scattering experiment in which the condensate atoms are diffracted by a standing light wave \citep{stenger}. The superradiant Rayleigh scattering from a BEC is the quantum analog of the collective atomic recoil laser (CARL) \citep{bonifacio}. Due to the absence of Doppler broadening in BEC, the atoms scatter a single laser photon and recoil with an extra momentum of $2 \hbar \vec{k}$ in the direction of the incident photon ($\vec{k}$ is the wave-vector of the incident photon). The amplitude of the matter-wave grating and the number of scattered photons are exponentially enhanced via the CARL instability \citep{Moore}. From the theoretical side, the semiclassical model of CARL was extended to include the quantum mechanical description of the center-of-mass motion of the atoms in a BEC \citep{Piovella}. The influence of atomic motion on the superradiant light acattering from a moving BEC was investigated theoretically and experimentally \citep{bonifacio2, fallani}. In an optical cavity due to the multiple reflections of the pump laser from the cavity mirrors, the interaction time of the light fields with the atoms can be enhanced by several orders of magnitude, which supports the amplification of the superradiant scattering process \citep{slama, motsch}.
Recently the field of cavity optomechanics has become an attractive research topic with a wide variety of systems ranging from gravitational wave detectors \citep{corbitt1, corbitt2}, nanomechanical cantilevers \citep{hohberger, gigan, arcizet, kleckner, favero, regal}, vibrating microtoroids\citep{carmon, schliesser}, membranes\citep{thompson} and Bose-Einstein condensate \citep{brennecke, murch, bhattacherjee09, bhattacherjee10, treulein, szirmai, singh}. A cavity opto-mechanical system, generally consists of an optical cavity with one movable end mirror. Such a system is utilized to cool a micromechanical resonator to its ground state by the pressure exerted by the cavity light field on the movable mirror.
The studies on cavity opto-mechanics of atoms show that sufficiently strong and coherent coupling would enable studies of atom-oscillator entanglement, quantum state transfer, and quantum control of mechanical force sensors. Due to coupling between the condensate wavefunction and the cantilever, mediated by the cavity photons, the cantilever displacement is expected to strongly influence the superfluid properties of the condensate \citep{bhattacherjee09}. Recently, coupled dynamics of a movable mirror and atoms trapped in the standing wave light field of a cavity were studied \citep{meiser}. It was shown that the dipole potential in which the atoms move is modified due to the back-action of the atoms and that the position of the atoms can become bistable. In this paper, we investigate the effects of a movable cavity mirror on the superradiant light scattering from a BEC in the quantum superradiant regime.

\section{The CARL-BEC model in cavity opto-mechanics}

\begin{figure}[h]
\hspace{-0.0cm}
\includegraphics [scale=0.6]{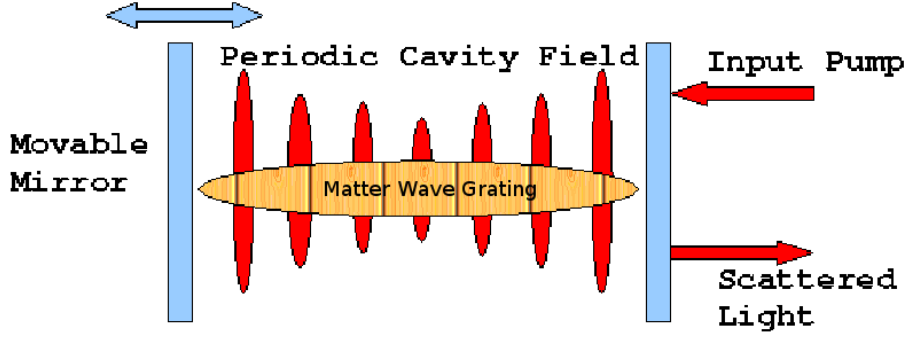}
\caption{Schematic representation of the system we are investigating here. The counter-propagating cavity light fields form a periodic optical grating which in turn forms a periodic matter wave grating. One of the cavity mirror is movable and completely reflecting. Pump light is incident from the fixed mirror which is partially reflecting. The scattered light that comes out of the fixed mirror is measured.  }
\label{f1}
\end{figure}

The system under consideration consists of noninteracting ultracold atoms interacting with two counterpropagating modes of an optical cavity. A particular system that we could consider is an elongated cigar shaped cloud of $N$ two-level $^{87} Rb$ atoms with mass $m$ and frequency $\omega_{a}$ of the $|F=1>$ $\rightarrow$ $|F'=2>$ transition. An external laser pump of frequency $\omega_{p}$ and amplitude $\eta$ is incident from one side mirror of the cavity as shown in Fig.1 and couples to the cavity field and drives the system. The induced resonance frequency shift of the cavity is assumed to be smaller than the longitudinal mode spacing, so that we restrict our model to a single longitudinal mode. An elongated cloud is created by keeping the frequency of the harmonic trap along the transverse direction larger than the frequency along the axial direction. The cavity mode is also coupled to a mechanical oscillator (one of the cavity mirror is allowed to move) with mechanical frequency $\Omega_{m}$. The strength of coupling between the cavity mode and the mechanical oscillator is $\epsilon$. The detuning of the pump laser frequency $\omega_{p}$ and the cavity mode frequency $\omega_{c}$ from the atomic transition frequency $\omega_{a}$ is assumed to be so large that the internal dynamics is continuously at a steady state. Keeping the population of the excited state at a negligible level. In this case then one can safely ignore spontaneous emission as well as two body dipole-dipole interactions. However, we still allow for two-photon virtual transitions in which the atomic internal states do not change, but due to recoil, the atom's center-of-mass motion will change. An atom which absorbs a photon from one mode and emits a photon into the counter propagating mode experiences a recoil kick equal to the difference of the momenta of the two photons (of the order of two optical momenta).
The two photon transitions couple different states of the atomic center-of-mass motion. The system described above can be modeled by the second quantized Hamiltonian

\begin{equation}\label{1}
H=H_{atom}+H_{photon}+H_{phonon}+H_{atom-photon}+H_{photon-phonon}+H_{pump},
\end{equation}

where, $H_{atom}$, $H_{photon}$ and $H_{phonon}$ give the free evolution of the atomic field, light field and the mechanical modes. $H_{atom-photon}$ and $H_{photon-phonon}$ describe the dipole coupling between the light field and the atomic field and the mechanical mode respectively. $H_{pump}$ is the pump field driving the system. The free one dimensional atomic Hamiltonian is given by

\begin{equation}\label{2}
H_{atom}=\int dz \left \{ \psi_{g}^{\dagger}(z) \left ( \frac{-\hbar^{2}}{2m } \frac{\partial^{2}}{\partial z^{2}} \right )\psi_{g}(z)+\psi_{e}^{\dagger}(z)\left (\frac{-\hbar^{2}}{2m } \frac{\partial^{2}}{\partial z^{2}}+\hbar \omega_{a} \right ) \psi_{e}(z) \right \}
\end{equation}

where $m$ is the atomic mass, $\omega_{a}$ is the atomic resonance frequency, $\psi_{e}(z)$ and $\psi_{g}(z)$ are the atomic field operators for the excited and ground state atoms respectively. The atomic field operators satisfy the bosonic equal time commutation relations $[\psi_{j}(z), \psi_{j'}^{\dagger}(z')]=\delta_{j,j'},\delta(z,z')$, and $[\psi_{j}(z),\psi_{j'}(z')]=[\psi_{j}^{\dagger}(z),\psi_{j'}^{\dagger}(z')]=0$, where $j,j'={e,g}$. The free evolution of the cavity mode is governed by the Hamiltonian

\begin{equation}\label{3}
H_{photon}=\hbar \omega_{c}a^{\dagger}a  ,
\end{equation}

where $\omega_{c}$ is the cavity frequency and $a$ and $a^{\dagger}$ are the cavity photon annihilation and creation operators respectively. The free evolution of the mechanical mode is given by the Hamiltonian

\begin{equation}\label{4}
H_{phonon}=\hbar \Omega_{m} a_{m}^{\dagger} a_{m},
\end{equation}

where $\Omega_{m}$ is the frequency of the mechanical mode and $a_{m}$ and $a_{m}^{\dagger}$ are the annihilation and creation operator for a mechanical mode respectively. The pump mode is governed by the Hamiltonian

\begin{equation}\label{5}
H_{pump}=-i \hbar \eta (a-a^{\dagger}),
\end{equation}

where $\eta$ is the strength of the pump field. The atomic fields and the counter propagating cavity light fields interact in the dipole approximation via the Hamiltonian

\begin{equation}\label{6}
H_{atom-photon}=-i\hbar g a \left \{\int dz \psi^{\dagger}_{e}(z) e^{ikz} \psi_{g}(z)+\int dz \psi^{\dagger}_{e}(z) e^{-ikz} \psi_{g}(z) \right \}  +H.C,
\end{equation}

where $g$ is the atom-light coupling constant. Here $k$ is the wavenumber of the cavity light mode. The counter-propagating cavity light modes are along the $\pm \hat{z}$ axis. Finally the Hamiltonian describing the coupling between the cavity photons and the mechanical mode is written as

\begin{equation}\label{7}
H_{photon-phonon}=\hbar \epsilon \Omega_{m} a^{\dagger}a (a_{m}+a^{\dagger}_{m}).
\end{equation}

The input pump laser populates the intracavity mode which couples to the moving mirror through the radiation pressure and the atoms through the dipole interaction.

Since the detuning $\Delta=\omega_{p}-\omega_{a}$ is large, spontaneous emission is negligible and we can adiabatically eliminate the excited state, using the Heisenberg equation of motion $\dot{\psi_{e}}=\frac{i}{\hbar}[H,\psi_{e}]$. This yields the Heisenberg equation of motion for the ground state field operator $\psi_{g}(z)$, the photon annihilation operator $a$, the position of the mechanical operator $x_{m}=(a_{m}+a_{m}^{\dagger})$ and the momentum of the mechanical oscillator $p_{m}=i(a_{m}^{\dagger}-a_{m})$.

\begin{equation}\label{8}
\frac{d}{dt}\psi_{g}(z)=\frac{i\hbar}{2m}\nabla^{2}\psi_{g}(z)-\frac{i2\hbar g^{2}}{\Delta}a'^{\dagger}a' \left \{1+\cos{2kz}\right \},
\end{equation}

\begin{equation}\label{9}
\frac{d}{dt}a'=i \Delta_{c} a'+\eta-\frac{i2g^{2}N}{\Delta}a'-\frac{12g^{2}}{\Delta}a' \int dz \psi_{g}^{\dagger}(z) \cos{2kz} \psi_{g}(z)-i\epsilon \Omega_{m} x_{m} a'-\kappa a',
\end{equation}

\begin{equation}\label{10}
\frac{d}{dt}x_{m}=\Omega_{m} p_{m},
\end{equation}

\begin{equation}\label{11}
\frac{d}{dt}p_{m}=-\Omega_{m}x_{m}-\Gamma_{m}p_{m}-2\epsilon \Omega_{m} a'^{\dagger}a',
\end{equation}

where $a'=a e^{i \omega_{p}t}$, $\Delta_{c}=\omega_{p}-\omega_{c}$ and $N=\int dz \psi_{g}^{\dagger}(z) \psi_{g}(z)$. In Eqn.\ref{8} the second term on the right side of the equality is the self consistent optical grating whose amplitude depends on time according to Eqn. \ref{9} . In Eqn. \ref{9} , the fourth term on the right side of equality is the self consistent matter wave grating and the fifth term is the radiation pressure coupling of the cavity mode with the mirror displacement.  The system we are considering is intrinsically open as the cavity field is damped by the photon leakage. This cavity photon decay rate is $\kappa$ and $\Gamma_{m}$ is the dissipation rate of the mechanical mode.

If the condensate is much longer than the radiation wavelength and the density is uniform, then periodic boundary condition can be assumed and the ground state wavefunction $\psi_{g}(z)$ can be written as

\begin{equation}\label{12}
\psi_{g}(z,t)= \sum_{n} C_{n}(t) e^{2inkz},
\end{equation}

where $e^{2inkz}$ are the momentum eigensfunctions with eigenvalues $p_{z}=n(2 \hbar k)$. This description of the atomic motion is equivalent to the assumption that the atoms in a BEC are delocalized inside the condensate and the momentum uncertainity is negligibly small. Using Eqn.\ref{12}, Eqn.\ref{8} and Eqn.\ref{9} reduce to the following set of ordinary differential equations,

\begin{equation}\label{13}
\frac{d}{dt}a'=i \left [\Delta_{c}-\epsilon \Omega_{m}x_{m} \right ]a'+\eta-\frac{i2g^{2}N}{\Delta}a'-\frac{i2g^{2}N}{\Delta}a'[C_{n}C_{n+1}^{*}+C_{n}^{*}C_{n-1}]-\kappa a',
\end{equation}

\begin{equation}\label{14}
\frac{d}{dt}C_{n}=-i4n^2\omega_{R}C_{n}-\frac{ig^{2}N}{\Delta}A[2C_{n}+C_{n-1}+C_{n+1}],
\end{equation}

\begin{equation}\label{15}
\frac{d}{dt}C_{n+1}=-i4(n+1)^2\omega_{R}C_{n+1}-\frac{ig^{2}N}{\Delta}A[2C_{n+1}+C_{n}],
\end{equation}

\begin{equation}\label{16}
\frac{d}{dt}C_{n-1}=-i4(n-1)^2\omega_{R}C_{n-1}-\frac{ig^{2}N}{\Delta}A[2C_{n-1}+C_{n}],
\end{equation}

where $A=a'^{\dagger}a'$ and we are assuming that only three momentum levels involved in the process are the initial level $n$, final levels $n+1$ and $n-1$. The levels $n+1$ and $n-1$ are the momentum side modes to the initial level $n$, generated by the two-photon transitions. We have introduced rescaled photon operators $a'\rightarrow\sqrt{N}a'$. In Eqn.(13), the first term on the right hand side is the renormalized cavity detuning. The cavity detuning is now changing with time due to $\epsilon \Omega_{m} x_{m}$.We note that when $\epsilon=0$ (stationary cavity), $\Delta_{c}=0$ is the Bragg condition of the scattering process, arising from energy and momentum conservation. We also note from Eqn.(13) that the mirror motion term has a dynamical dispersive effect on the Bragg resonances, proportional to the mirror displacement $x_{m}$. In the linear regime, when $a$ is still small and $x_{m}=x_{o}$ (some initial constant value), the Bragg condition is $\Delta_{c}=\epsilon \Omega_{m} x_{o}$, which is just a shift of the Bragg resonance.

Defining coherences, $S_{+}=C_{n}C_{n+1}^{*}$, $S_{-}=C_{n}C_{n-1}^{*}$, $S_{-+}=C_{n-1}C_{n+1}^{*}$, the population differences $W_{1}=|C_{n}|^{2}-|C_{n+1}|^{2}$ and $W_{2}=|C_{n-1}|^{2}-|C_{n}|^{2}$, we have the following additional equations of motion.

\begin{equation}\label{17}
\frac{d}{dt}S_{+}=i4(1+2n)\omega_{R}S_{+}+\frac{ig^{2}N}{\Delta}A[W_{1}-S_{-+}]-\gamma_{+}S_{+},
\end{equation}

\begin{equation}\label{18}
\frac{d}{dt}S_{-}=-i4(1-2n)\omega_{R}S_{-}+\frac{ig^{2}N}{\Delta}A[W_{2}+S_{-+}]-\gamma_{-}S_{-},
\end{equation}

\begin{equation}\label{19}
\frac{d}{dt}W_{1}=-\frac{2g^{2}N}{\Delta}A[2 S_{+}^{i}+S_{-}^{i}],
\end{equation}

\begin{equation}\label{20}
\frac{d}{dt}W_{2}=\frac{2g^{2}N}{\Delta}A[2 S_{-}^{i}+S_{+}^{i}],
\end{equation}

\begin{equation}\label{21}
\frac{d}{dt}S_{-+}^{r}=-16n \omega_{R}S_{-+}^{i}+\frac{g^{2}N}{\Delta}A (S_{+}^{i}-S_{-}^{i})-\gamma_{-+}S_{-+}^{r},
\end{equation}

\begin{equation}\label{22}
\frac{d}{dt}S_{-+}^{i}=16n \omega_{R}S_{-+}^{r}-\frac{g^{2}N}{\Delta}A (S_{+}^{r}-S_{-}^{r})-\gamma_{-+}S_{-+}^{i},
\end{equation}

where, superscripts $i$ and $r$ indicate the imaginary and real parts of the corresponding operators. $\gamma_{i}(i=+,-,-+)$ are the decoherence rates due to Doppler broadening, inhomogeneous effects and phase diffusion. The phase diffusion mechanism depends on the frequency detuning between the incident and scattered radiation beams and on the initial momentum of the condensate $p_{o}=n(2\hbar k)$. Eqns. \ref{13}-\ref{22}, are then formally equivalent to the well known Maxwell-Bloch equations for a $3$-level system.

\begin{figure}[h]
\hspace{-0.0cm}
\includegraphics [scale=0.7]{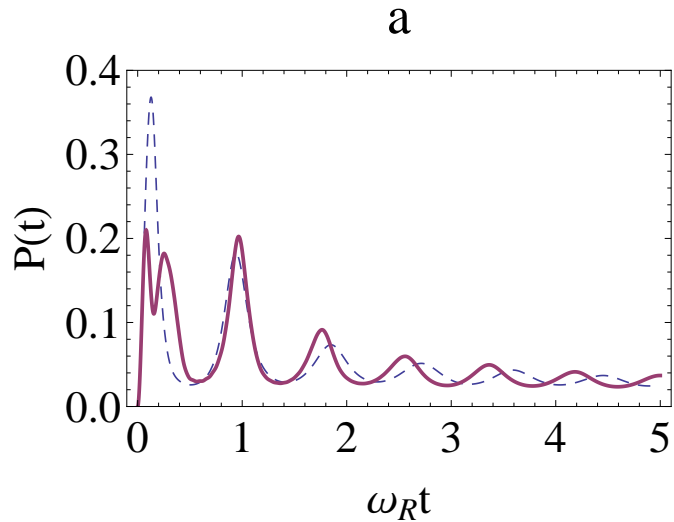}
\caption{Time signal of the scattered light power $P(t)=a'^{\dagger} a'$ in the classical superradiant regime. All frequencies are in units of the recoil frequency $\omega_{R}$ Parameters used are: $\Delta_{c}=5$, $\frac{g^{2}N}{\Delta}=20$, $\eta=10$, $n=0$, $\kappa=10$, $\Gamma_{m}=1$, $\Omega_{m}=100$, $\gamma_{+}=\gamma_{-}=\gamma_{-,+}=0.1$, $\epsilon=0.1$(dashed line) and $\epsilon=0.7$ (solid line).  }
\label{f2}
\end{figure}

Before we proceed ahead with our analysis of the above equations, we identify two regimes of interest. In the quantum superradiant regime $\frac{g^{2}N}{\Delta}<\kappa$, and each atom scatters only a single photon coherently and the condensate momentum changes by $2\hbar k$. In the semiclassical superradiant regime $\frac{g^{2}N}{\Delta}>\kappa$, the momentum gained by the atoms scattering photons is much larger than $\hbar k$.
The collective dynamics begin as soon as the pump power builds up in the ring cavity. We study this dynamics via the evolution of the power $P(t)=a'^{\dagger}a'$. In the semi-classical regime, the time signal of the probe light shows characteristic maxima and minima as shown in Fig.2. Here, we have taken the BEC to be at rest initially ($n=0$). This behaviour can be explained as follows: The atoms occupy an initial momentum eigenstate and are coupled to the final momentum state by the two-photon transition. The time evolution is like the usual Rabi oscillation. This coherent dynamics leads to the build up of an atomic density in each state and minimum when all atoms are in the initial or final state. the scattered light is proportional to the contrast of the atomic density grating. Maxima in $P(t)$ is observed whenever the momentum state of the atoms change. The temporal decay of the atomic density grating due to various decoherence mechanisms. In Fig.2, we show two curves for two different values of the mirror-photon coupling strength, $\epsilon=0.1$ (dashed curve), $\epsilon=0.7$(solid curve). In the semi-classical regime, the influence of the micro-motion of the mirror is almost negligible. The temporal evolution of the light power $P(t)$ shown in Fig.2 is very much similar to that measured experimentally in a stationary cavity \citep{slama}. Eventhough, the above quantum Maxwell-Bloch equations are strictly valid in the $g^{2}N/\Delta<\kappa$ regime, we still see that it reproduces qualitatively the experimental superradiant spectrum well in the $g^{2}N/\Delta>\kappa$ regime also. In the $g^{2}N/\Delta>\kappa$ regime, the atomic motion is treated classically. In the following, we only study the quantum superradiant regime.

\begin{figure}[h]
\hspace{-0.0cm}
\begin{tabular}{cc}
\includegraphics [scale=0.65]{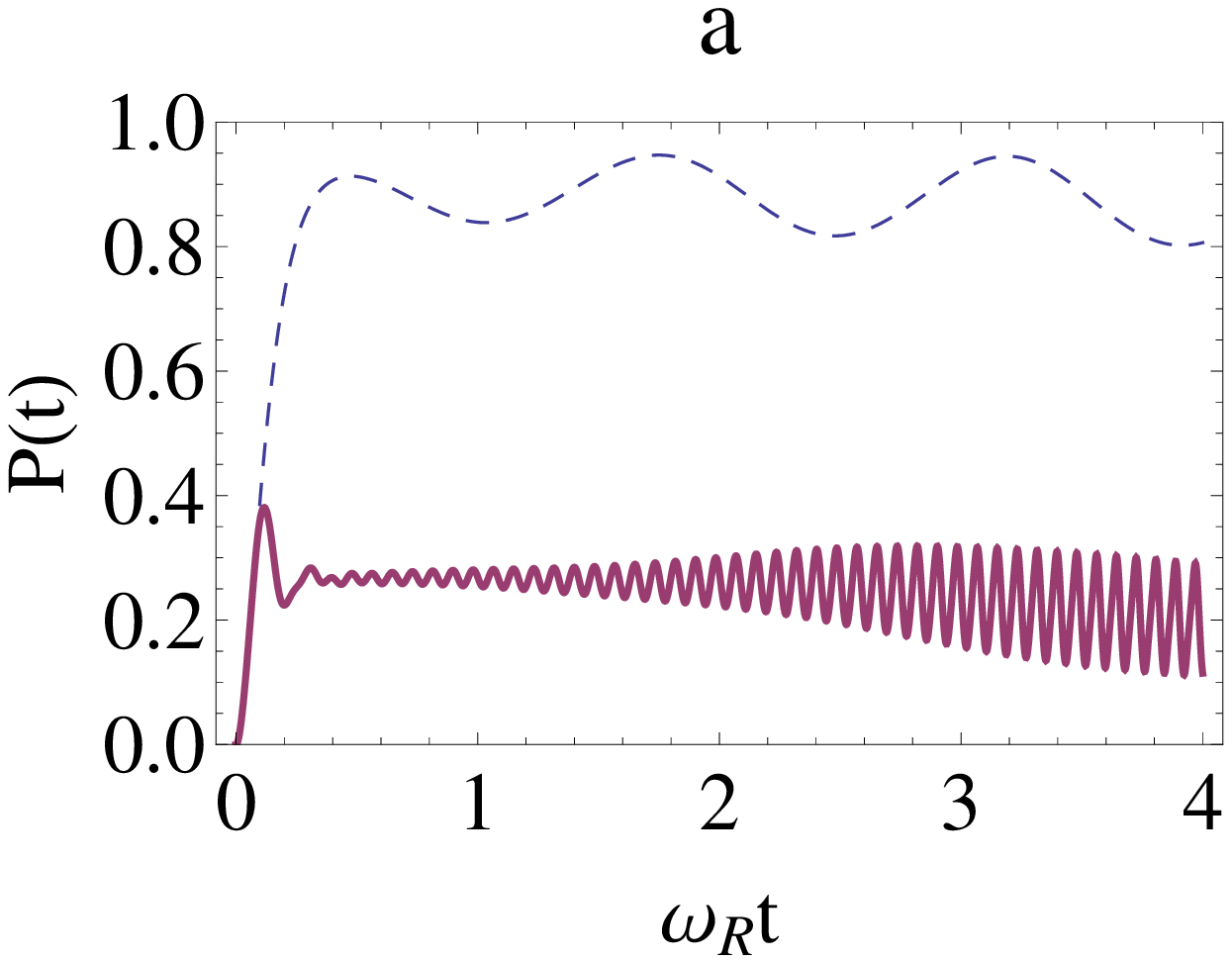}& \includegraphics [scale=0.65] {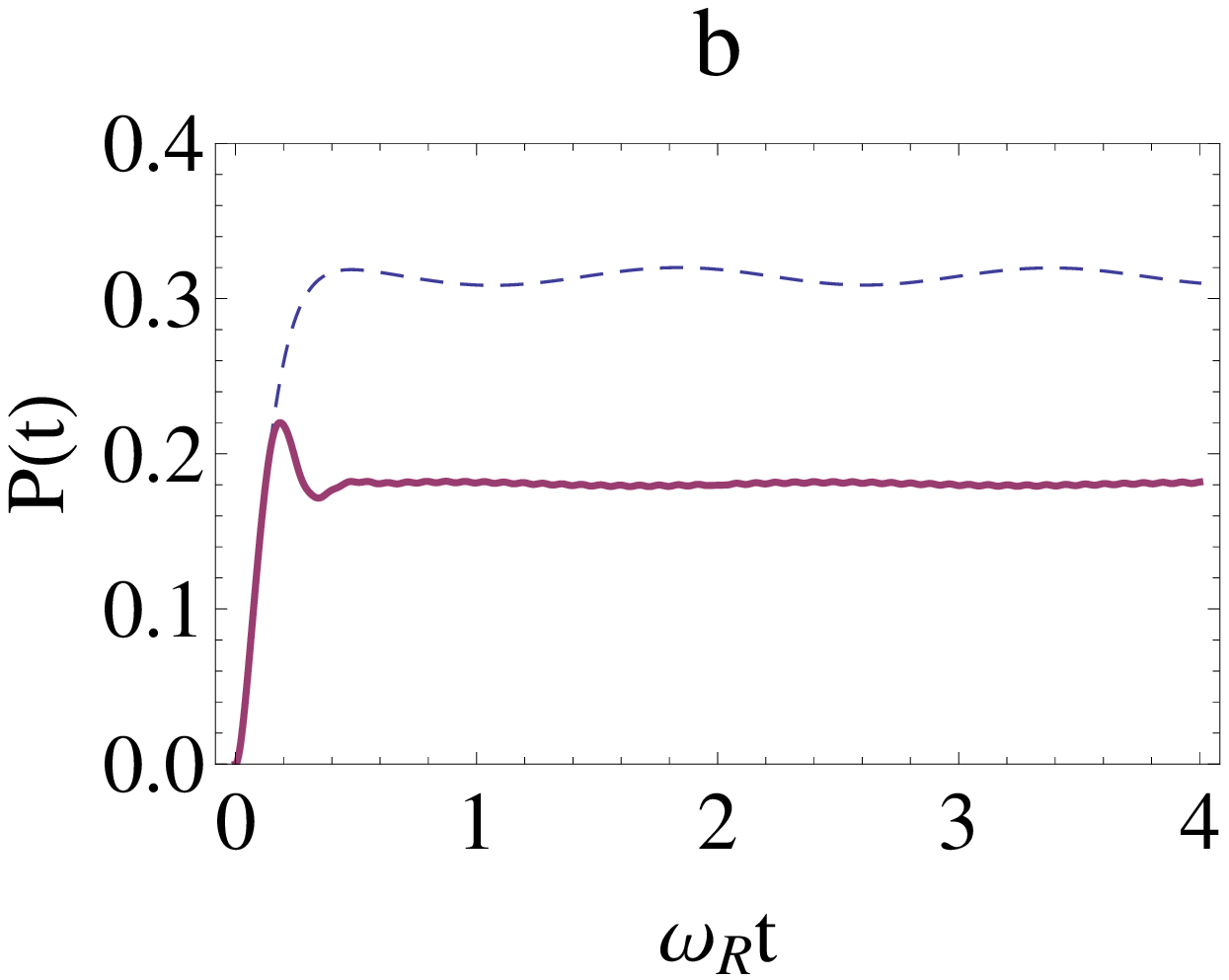}\\
 \end{tabular}
\caption{ Time signal of the scattered light power $P(t)=a'^{\dagger} a'$ in the quantum superradiant regime. All frequencies are in units of the recoil frequency $\omega_{R}$ Parameters used are: $\Delta_{c}=5$, $\frac{g^{2}N}{\Delta}=1$,  $n=0$, $\kappa=10$, $\Gamma_{m}=1$, $\Omega_{m}=100$, $\gamma_{+}=\gamma_{-}=\gamma_{-,+}=0.1$, $\epsilon=0.1$(dashed line) and $\epsilon=0.7$ (solid line). For plot (a): $\eta=10$ and plot (b): $\eta=6$.}
\label{f3}
\end{figure}

We now investigate the transition from the Bragg scattering regime to the quantum superradiant regime. We start with a BEC at rest. As shown in Fig.3, we observe the transition between the two regimes by varying the pump amplitude. The plots in Fig.3(a) corresponds to $\eta/\omega_{R}=10$ ($\epsilon=0.1$ (dashed curve)and $\epsilon=0.7$ (solid curve)). Note that from Eqn. \ref{13} that when $2g^2N/\Delta <<1$, each time $\Delta_{c}-\epsilon \Omega_{m} x_{m}=0$, the light spectrum proportional to $a'^{\dagger}a'$ will show a peak. For large times, $t >> \kappa^{-1}$,

\begin{equation}
a'=\frac{\eta}{\kappa-i \left \{\Delta_{c}-\epsilon \Omega_{m} x_{m}-\frac{2 g^2 N}{\Delta} [1+S_{+}+S_{-}] \right \}}
\end{equation}

In this case , we observe the typical Bragg resonances due to two-photon transitions between the momentum modes. Reducing the pump intensity to $\eta/\omega_{R}=6$ in Fig.3(b), the Bragg oscillations vanishes. A reduction in the pump power leads to a decrease of the contrast of the optical standing wave resulting from the interference of the counter-propagating light beams. This result is in accordance with previous theoretical result \citep{fallani}.

\begin{figure}[h]
\hspace{-0.0cm}
\includegraphics [scale=0.7] {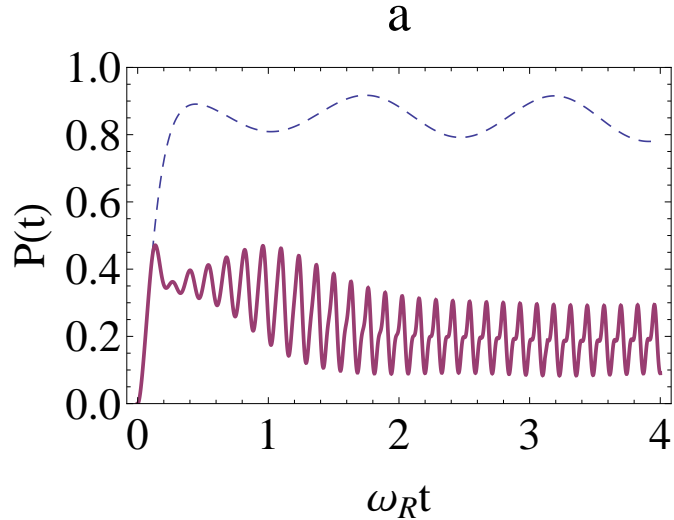}\\
\caption{ Influence of the mechanical frequency on the time signal of the scattered light power $P(t)=a'^{\dagger} a'$ in the quantum superradiant regime. All frequencies are in units of the recoil frequency $\omega_{R}$ Parameters used are: $\Delta_{c}=5$, $\frac{g^{2}N}{\Delta}=1$, $\eta=10$, $n=0$, $\kappa=10$, $\Gamma_{m}=1$, $\Omega_{m}=50$, $\gamma_{+}=\gamma_{-}=\gamma_{-,+}=0.1$, $\epsilon=0.1$(dashed line) and $\epsilon=0.7$ (solid line). Compare this figure with Fig 2(a).}
\label{f4}
\end{figure}

The influence of decreasing the mechanical frequency is shown in Fig. 4. We note that the effect of decreasing the mechanical frequency is to broaden the superradiant signals (with a corresponding increase in the duration between two signals) together with an increase in the peak intensity of the superradiant signal. This is because as we are decreasing the mechanical frequency, the dispersive effect reduces. On decreasing the mechanical frequency further, the solid curve approaches the dashed curve. In the limit of extremely small mechanical frequency ($\Omega_{m}$) or extremely low opto-mechanical coupling ($\epsilon$),the density grating contrast almost remains the same. Since the scattered light is proportional to the density grating contrast, this leads to observation that the maxima and the minima in the scattered light power remains the same for a substantial duration with a very slow decay of the peak power. The time difference between the two Bragg peaks also corresponds to the typical time scale, on which the atomic momentum distribution is shuffled between different momentum states.

\begin{figure}[t]
\hspace{-0.0cm}
\begin{tabular}{cc}
\includegraphics [scale=0.65]{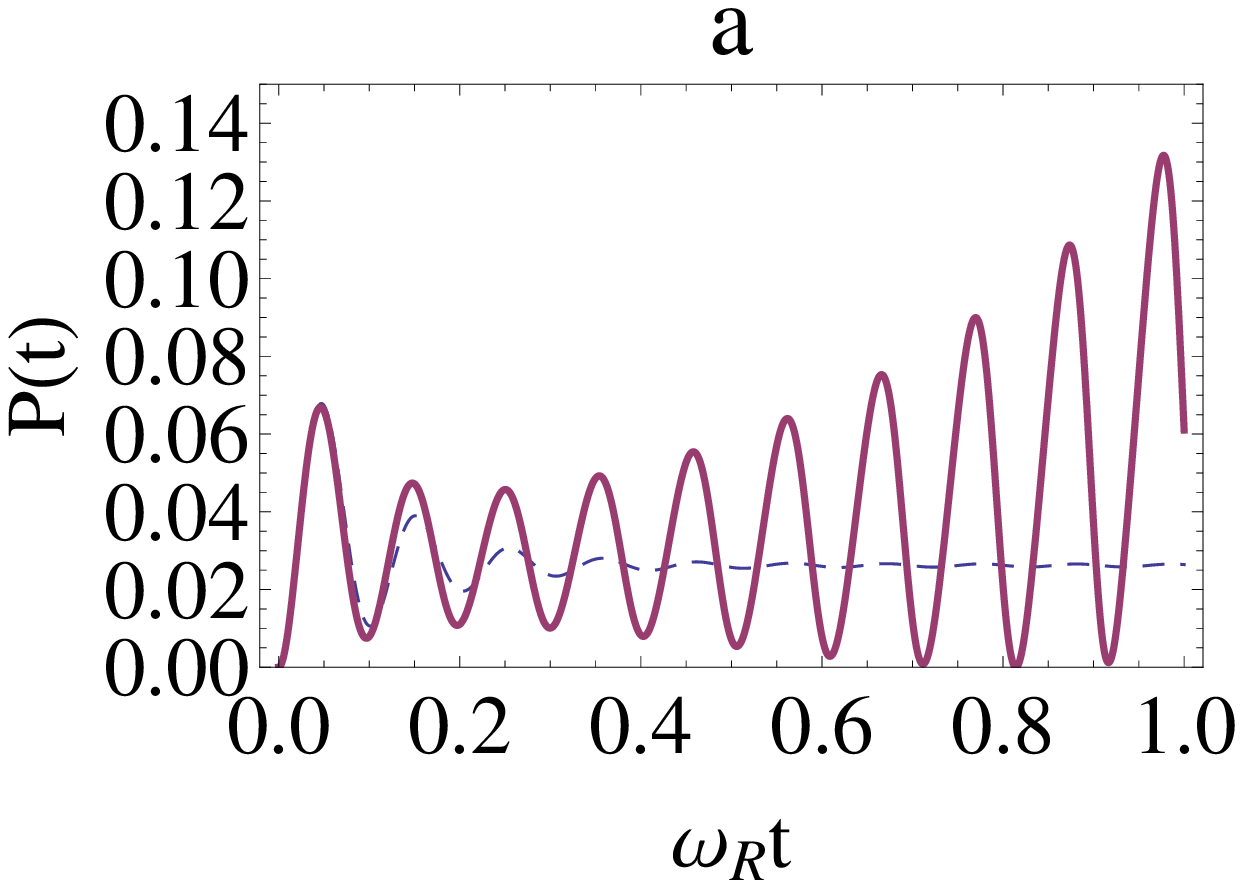}& \includegraphics [scale=0.65] {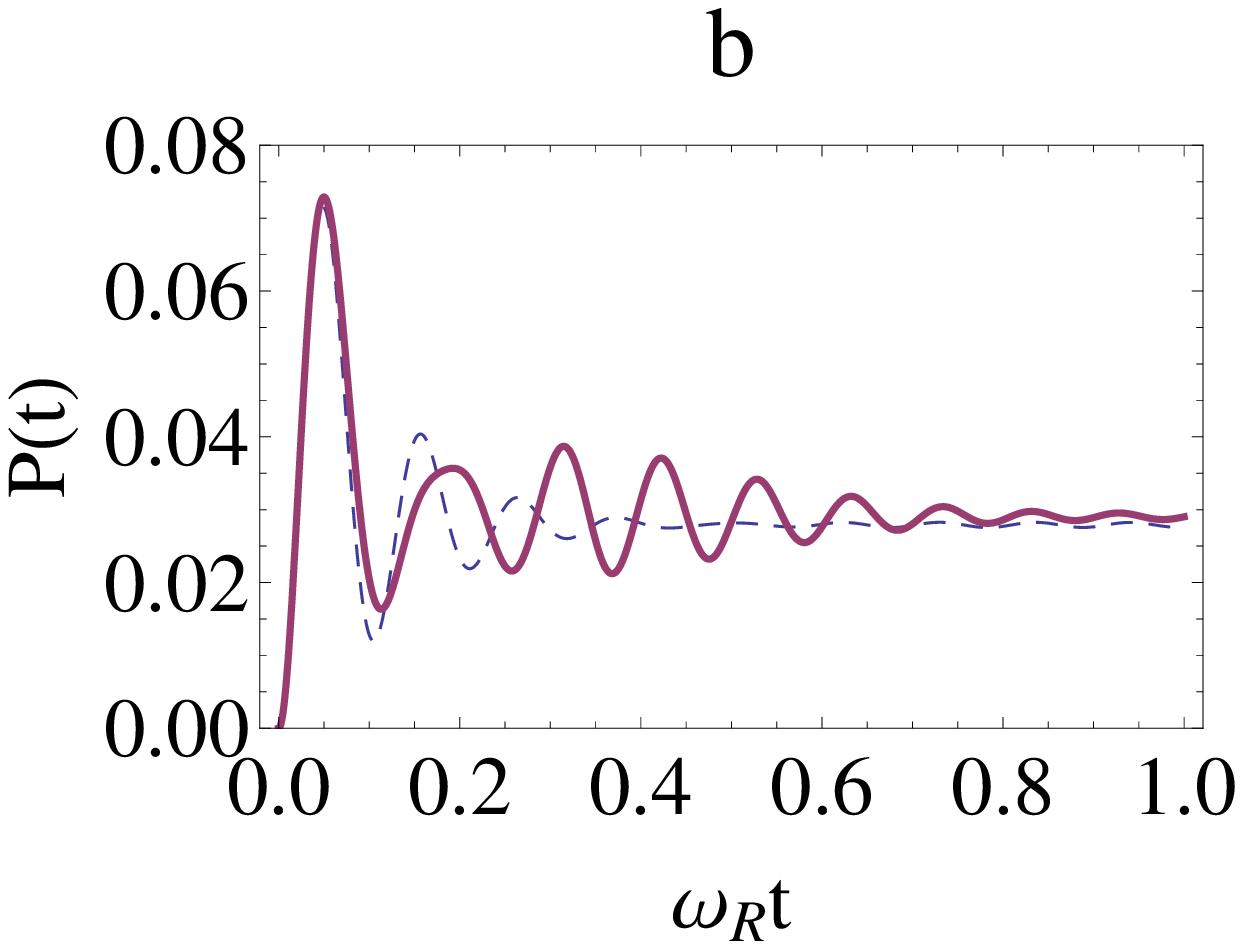}\\
 \end{tabular}
\caption{ Influence of the cavity detuning on the time signal of the scattered light power $P(t)=a'^{\dagger} a'$ in the quantum superradiant regime. All frequencies are in units of the recoil frequency $\omega_{R}$ Parameters used are: $\frac{g^{2}N}{\Delta}=1$, $\eta=10$, $n=0$, $\kappa=10$, $\Gamma_{m}=1$, $\Omega_{m}=50$, $\gamma_{+}=\gamma_{-}=\gamma_{-,+}=0.1$, $\epsilon=0.1$(dashed line) and $\epsilon=0.7$ (solid line). For plot (a): $\Delta_{c}=-\Omega_{m}$ and plot (b): $\Delta_{c}=\Omega_{m}$.}
\label{f5}
\end{figure}

In order to coherently control the superradiant process, we immediately identify a new handle which is inherently present in the current system, i.e the process by which the mechanical mode of the moving mirror can be cooled. It is well known that a red detuning of the pump laser frequency $\omega_{p}$ from the cavity mode $\omega_{c}$ by an amount $\Omega_{m}$ will ensure that the desired anti-Strokes process (required for cooling the mirror to its ground state) is resonantly enhanced while the deleterious Strokes process, being off-resonant, is suppressed \citep{rae}. Consequently, we expect that when $\Delta_{c}= -\Omega_{m}$, the superradiant process will be amplified since the corresponding mechanical motion of the mirror will be suppressed due to the opto-mechanical cooling process by which energy from the mechanical mode is pumped into the photons. On the other hand when $\Delta_{c}= +\Omega_{m}$, the superradiant process is suppressed while the mechanical motion of the mirror is amplified due to energy transfer from the photons to the mechanical mode. This is exactly what we found in the resolved side band (RSB) regime ($\Delta_{c}=-\Omega$ and $\Omega_{m}>\kappa$ ) and the results are depicted in Fig.5. We note that the intensity of the scattered light is considerably reduced in the RSB regime. Fig.5(a) is for the case $\Delta_{c}= -\Omega_{m}$ while Fig5(b) is for $\Delta_{c}= +\Omega_{m}$. Note that this observation is only for strong mirror-photon coupling. We also found (figures not shown) that the amplitude of the mechanical mode $x_{m}$ was amplified for $\Delta_{c}= +\Omega_{m}$ and was suppressed for $\Delta_{c}= -\Omega_{m}$.

\begin{figure}[t]
\hspace{-0.0cm}
\includegraphics [scale=0.7]{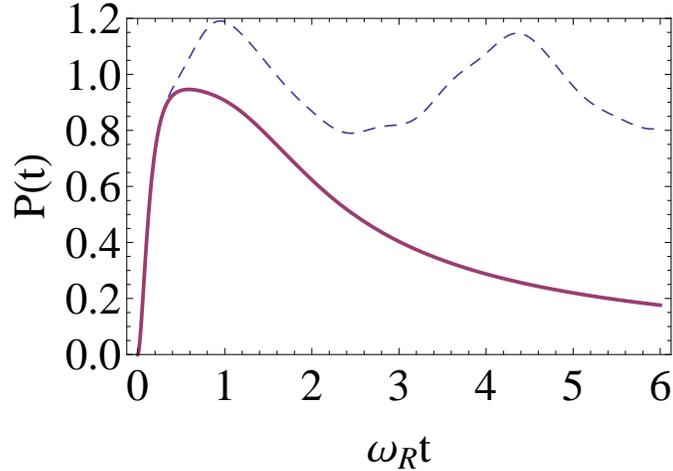}
\caption{ Influence of the initial atomic momentum $n$ on the time signal of the scattered light power $P(t)=a'^{\dagger} a'$ in the quantum superradiant regime. All frequencies are in units of the recoil frequency $\omega_{R}$ Parameters used are: $\Delta_{c}=5$, $\frac{g^{2}N}{\Delta}=1$, $\eta=10$, $\epsilon=0.1$ $\kappa=10$, $\Gamma_{m}=1$, $\Omega_{m}=100$, $\gamma_{+}=\gamma_{-}=\gamma_{-,+}=0.1$, $n=-0.5$(dashed line) and $n=0.5$ (solid line). Compare this figure with the dashed plot of Fig 2(a). }
\label{f6}
\end{figure}

The efficiency of the superradiant process, driven by the coherent coupling between the matter waves, optical field and the phonons, strongly depends on the coherence of the matter-wave grating. The matter-wave grating has a finite lifetime . After this time, the matter-wave grating decays and the superradiant process stops. The decay of the matter-wave grating depends on the initial momentum state of the condensate. In Fig.6, we show two plots, one for $n=0.5$ (solid curve) and for $n=-0.5$ (dashed). As evident, the light power for $n=0.5$ drops fast without appreciable Bragg scattering peaks. On the other hand for $n=-0.5$, the output light power has spectrum intermediate between that for $n=0$ (dashed curve of Fig.3(a)) and for $n=0.5$. In the RSB regime, we found that the initial momentum did not influence the output spectrum anymore. This is due to the fact that at large detuning $\Delta_{c}$, the influence of the term $\epsilon \Omega_{m} x_{m}$ is considerably reduced and since the influence of the initial condensate momentum on the scattered light comes in through this term, the power spectrum $<a'^{\dagger}a'>$ is effectively independent of the initial momentum.
To demonstrate that the dynamics investigated here are within experimental reach, we discuss the experimental parameters from \citep{brennecke,murch,schliesser2}: A BEC of typically $10^{5}$ $^{87}Rb$ atoms is coupled to the light field of an optical ultra high-finesse Fabry-Perot cavity. The atom-field coupling $g_{0}=2 \pi \times 10.9 Mhz$ \citep{brennecke} ( $2 \pi \times 14.4 $ \citep{murch}) is greater than the decay rate of the intracavity field $\kappa=2 \pi \times 1.3 Mhz$ \citep{brennecke} ($2 \pi \times 0.66 Mhz$ \citep{murch}). The temperature of the ultracold atomic gas $T<<\hbar \kappa/k_{B}$ so that the coherent amplification or damping of the atomic motion is neglected. Typically atom-pump detuning is $2 \pi \times 32 Ghz$. From \citep{schliesser2},the mechanical frequency $\Omega_{m}=2 \pi \times 73.5 Mhz$ and $\Gamma_{m}=2 \pi \times 1.3 Khz$. The coupling rate $g_{m}=2 \pi \times 2.0 Mhz$.
The energy of the cavity mode decreases due to the photon loss through the cavity mirrors, which leads to a reduced atom-field coupling. Photon loss can be minimized by using high-Q cavities. Our proposed detection scheme relies crucially on the fact that coherent dynamics dominate over the losses. It is important that the characteristic time-scales of coherent dynamics are significantly faster than those associated with losses (the decay rate of state-of-art optical cavities is typically 17 kHz \citep{Klinner06}).

\section{Conclusions}
In conclusion, we have analyzed how the superradiant light scattering from an elongated Bose-Einstein condensate in an optical cavity in the quantum regime is modified by the micro-motion of one of the cavity mirrors. Due to the coupling between the cavity photons and the mirror, the mirror motion has a dispersive effect on the Bragg resonance. The resulting Bragg resonance becomes time dependent. The spectrum of the scattered light shows Bragg oscillations (modified by the mirror motion) superimposed on the usual superradiant spectrum. The mechanical frequency of the mirror and the cavity-pump detuning is found to be a new handle to coherently control and manipulate the superradiant light scattering process. In particular, we found that in the resolved side band regime of the mirror cooling, the superradiant process is enhanced.

\section{Acknowledgements}

The author acknowledges a fellowship from the German Academic Exchange Service (DAAD) and is also grateful to Dr. Axel Pelster for facilities to carry out part of this work at the Freie University, Berlin.


\begin{thebibliography}{plain}

\bibitem{Inouye}
S. Inouye, et al., Science, {\textbf{285}}, 571, (1999).
%
\bibitem{Schneble}
D. Schneble et al., Science, {\textbf{300}},475, (2003).
%
\bibitem{kozuma}
M. Kozuma et al., Science, {\textbf{286}}, 2309, (1999).
%
\bibitem{stenger}
J. Stenger et al., Phys. Rev. Lett., {\textbf{82}}, 4569, (1999).
%
\bibitem{bonifacio}
R. Bonifacio and L. De Salvo Souza, Nucl. Instrum. and Meth. Phys. Res.A {\textbf{341}}, 360, (1994).
%
\bibitem{Moore}
M. G. Moore and P. Meystre, Phys. Rev. A., {\textbf{60}}, 1491, (1999).
%
\bibitem{Piovella}
N. Piovella, M. Gatelli and R. Bonifacio, Opt. Comm., {\textbf{194}}, 167, (2001).
%
\bibitem{bonifacio2}
R. Bonifacio et al., Opt. Comm. {\textbf{233}}, 155, (2004).
%
\bibitem{fallani}
L. Fallani et al., Phys. Rev. A, {\textbf{71}}, 033612, (2005).
%
\bibitem{slama}
S. Slama et al., Phys. Rev. A, {\textbf{75}}, 063602, (2007).
%
\bibitem{motsch}
M. Motsch et al., New J. Phys., {\textbf{12}}, 063022, (2010).
%
\bibitem{corbitt1}
T. Corbitt and N. Mavalvala, J. Opt. B: Quantum Semi-class. Opt. {\textbf{6}}, S675 (2004).
%
\bibitem{corbitt2}
T. Corbitt et al., Phy. Rev. Letts. {\textbf{98}}, 150802 (2007).
%
\bibitem{hohberger}
C. H\"{o}hberger-Metzger and K. Karrai, Nature {\textbf{432}}, 1002 (2004).
%
\bibitem{gigan}
S. Gigan et al., Nature {\textbf{444}}, 67 (2006).
%
\bibitem{arcizet}
O. Arcizet et al., Nature {\textbf{444}}, 71 (2006).
%
\bibitem{kleckner}
D. Kleckner and D. Bouwmeester, Nature {\textbf{444}}, 75 (2006).
%
\bibitem{favero}
I. Favero et al., Appl. Phys. Lett. {\textbf{90}}, 104101 (2007).
%
\bibitem{regal}
C. Regal, J. D. Teufel and K. Lehnert, Nature Physics, {\textbf{4}}, 555 (2008).
%
\bibitem{carmon}
T. Carmon et al., Phys. Rev. Letts. {\textbf{94}}, 223902 (2005).
%
\bibitem{schliesser}
A. Schliesser et al., Phys. Rev. Letts. {\textbf{97}}, 243905 (2006).
%
\bibitem{thompson}
J. D. Thompson et al., Nature {\textbf{452}}, 72 (2008).
%
\bibitem{brennecke}
F. Brennecke et al., Science {\textbf{322}}, 235 (2008).
%
\bibitem{murch}
K. W. Murch et al., Nature Physics {\textbf{4}}, 561 (2008).
%
\bibitem{bhattacherjee09}
A. Bhattacherjee, Phys. Rev. A, {\textbf{80}}, 043607 (2009).
%
\bibitem{bhattacherjee10}
A. Bhattacherjee, J. Phys. B., {\textbf{43}},205301, (2010).
%
\bibitem{treulein}
P. Treutlein et al., Phys. Rev. Letts., {\textbf{99}}, 140403 (2007).
%
\bibitem{szirmai}
G. Szirmai, D. Nagy and P. Domokos, Phys. Rev. A, {\textbf{81}}, 043639 (2010).
%
\bibitem{singh}
S. Singh, M. Bhattacharya, O. Dutta, and P. Meystre,  Phys. Rev. Lett. {\textbf {101}}, 263603 (2008), C. Genes, D. Vitali, and P. Tombesi,  Phys. Rev. A {\textbf {77}}, 050307 (2008), H. Ian, Z. R. Gong, Y. Liu, C. P. Sun, and F. Nori,  Phys. Rev. A {\textbf {78}}, 013824 (2008), A. A. Geraci and J. Kitching, Phys. Rev. A {\textbf {80}}, 032317 (2009), K. Hammerer, M. Wallquist, C. Genes, M. Ludwig, F. Marquardt, P. Treutlein, P. Zoller, J. Ye, and H. J. Kimble, Phys. Rev. Lett. {\textbf {103}}, 063005 (2009). S.K. Steinke and P. Meystre, arXiv:1103.6078. D. Hunger et al., arXiv:1103.1820, Bib Chen, Cheng Jiang and Ka-Di Zhu, Phys. Rev.A {\textbf{83}}, 055803 (2011).
%
\bibitem{meiser}
D. Meiser, and P. Meystre, Phys. Rev. A., {\textbf{73}}, 033417 (2006).
%
\bibitem{rae}
I. Wilson-Rae, N. Nooshi, W. Zwerger and T. J. Kippenberg, Phys. Rev. Letts., {\textbf{99}}, 093901, (2007).
%
\bibitem{schliesser2}
A. Schliesser et al., Nature Phys. ,{\textbf{4}}, 415 (2008).
%
\bibitem{Klinner06}
J. Klinner, M. Lindholdt, B. Nagorny and A. Hemmerich, Phys. Rev. Letts. ,{\textbf{96}}, 023002 (2006).
%
\end{thebibliography}
\end{document}